\documentclass[journal=jpclcd,layout=twocolumn,manuscript=letter]{achemso}
\setkeys{acs}{maxauthors=8, etalmode=truncate}

\usepackage{graphicx}
\usepackage{amssymb}
\usepackage{booktabs}
\usepackage{array}
\usepackage{mathtools}
\usepackage{amssymb}
\usepackage{upgreek}
\usepackage{float}
\usepackage{subfig}
\usepackage{ragged2e} 
\usepackage[normalem]{ulem}
\usepackage{color}
\usepackage[range-phrase=\text{--}, range-units=single]{siunitx}
\DeclareSIUnit\angstrom{\text{Å}}
\usepackage{comment}
\usepackage{multirow}
\usepackage{setspace}
\usepackage{enumitem}
\setlist{nosep}

\usepackage{ifthen}
\usepackage[section]{placeins}
\usepackage{hyperref}
\definecolor{jade}{rgb}{0.0, 0.66, 0.42}
\definecolor{lapislazuli}{rgb}{0.15, 0.38, 0.61}
\hypersetup{
    colorlinks=true,
    linkcolor=lapislazuli,
    citecolor=jade,
    }
\usepackage[multidot]{grffile}
\graphicspath{{Figures/}{SIFigures/}}

\newcommand{\dir}{./Bib}

\title{Lost in Projection? Gaussian Filtering Recovers Hidden Conformational States}
\author{Sofia Sartore}
\author{Daniel Nagel}
\author{Georg Diez}
\author{Gerhard Stock}
\email{stock@physik.uni-freiburg.de}
\affiliation{Biomolecular Dynamics, Institute of Physics,
   University of Freiburg, 79104 Freiburg, Germany}
\date{\today}
\begin{document}

\begin{tocentry}
\includegraphics{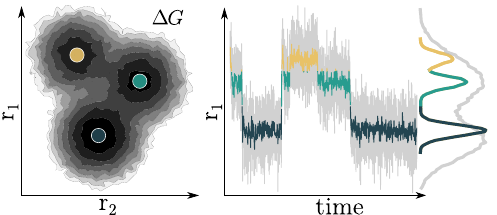}
\end{tocentry}

\begin{abstract} 
  \baselineskip5mm
To interpret molecular dynamics (MD) simulations, it is common
practice to reduce the dimensionality of the molecular coordinates to
a low-dimensional collective variable $x$. Projecting the
high-dimensional MD data onto $x$ yields a free energy landscape
$\Delta G(x)$, which highlights low-energy regions corresponding to
conformational states. The accurate definition of these states,
however, is often impeded by projection artifacts, resulting in
artificially shortened state lifetimes or even the complete
disappearance of states from the analysis. As demonstrated for a
two-dimensional toy model, Gaussian low-pass filtering of the
high-dimensional MD coordinates can restore the underlying free energy
landscape, allowing to recover previously hidden states. When applied to
an all-atom folding trajectory of HP35, the number of microstates
increases by an order of magnitude, which leads to metastable states
that are long-lived and much better defined structurally, even
compared to dynamically cored state trajectories.
\end{abstract}

%
%
\baselineskip5mm

Complex dynamical systems, like solvated proteins, are characterized
by dynamics that span multiple timescales. The output of
molecular dynamics (MD) simulations provides the time evolution of the
atomic coordinates, capturing fast local fluctuations and slower
global conformational rearrangements \cite{berendsen_simulating_2007}. 
Since the latter usually constitute the processes of interest, the analysis of MD simulations involves separating fast and slow motions. This is often
done by identifying metastable conformational states and modeling the
dynamics of interest by a series of memoryless transitions between
these states, yielding a Markov state model (MSM)
\cite{chodera_obtaining_2006, buchete_coarse_2008,
  bowman_progress_2009, prinz_markov_2011, bowman_introduction_2014,
  wang_constructing_2018}. This approach is widely used, since it
holds the promise of predicting long-term dynamics from short
trajectories, and because its use is facilitated by workflow implementations in
software packages like PyEmma\cite{scherer_pyemma_2015},
MSMBuilder,\cite{beauchamp_msmbuilder2_2011} and
msmhelper\cite{nagel_msmhelper_2023}.

Building meaningful MSMs from MD data, however, presents the challenge
of correctly identifying metastable states. The
high-dimensional space of atomic coordinates compromises this
task at the very start, because the sparsity of the data distribution hinders statistical
robustness. This issue can in principle be overcome by the assumption that protein
dynamics reside on low-dimensional manifolds.  \cite{hegger_how_2007,
  facco_estimating_2017}. All dimensionality reduction techniques,
from straightforward linear principal component
analysis\cite{amadei_essential_1993} to sophisticated deep-learning autoencoder strategies \cite{wang_machine_2020, glielmo_unsupervised_2021}
exploit this manifold hypothesis in order to project the initial coordinates onto a low-dimensional space of collective variables
and effectively capture essential dynamics. The reduced-dimensional energy landscape of such projected coordinates is then scanned with a clustering algorithm that identifies the minima corresponding to metastable states. Density-based methods\cite{keller_comparing_2010, rodriguez_clustering_2014, sittel_robust_2016, song_adaptive_2017} are a valuable alternative to the widely used geometric clusterings (such as $k$-means\cite{jain_data_2010}) in MSM construction, as they perform better in defining states at the barriers\cite{sittel_perspective_2018}. 
However, despite many advances in dimensionality reduction and clustering techniques, projecting high-dimensional data onto a lower-dimensional space often leads to misclassifications of data points, especially in sparsely populated transition regions. This hampers a proper definition of the states and their metastability, 
causing fast intrastate fluctuations to be interpreted as interstate transitions, which leads to an underestimation of the timescales of the processes under study.
Even worse, it may also lead to the complete disappearance of states from the analysis.

A simple way of correcting these projection artifacts was introduced
with the concept of coring.\cite{buchete_coarse_2008} From a geometrical perspective, a core is defined as the region surrounding the center of a state. The idea is that the
system should reach that region for a transition to a new
state to be considered completed, thus resulting in a modified state
trajectory.\cite{schutte_markov_2011,
  lemke_density-based_2016}. However, as the definition of geometrical
cores is not always trivial, especially in the case of high
dimensional coordinates, dynamical variants of coring were also
introduced. These are generally based on the requirement that the
system spends a minimum amount of time---the coring time $t_{\text{cor}}$---in the new state after
transitioning\cite{jain_hierarchical_2014, nagel_dynamical_2019}.  
Hence, $t_{\text{cor}}$ poses a limit for
the time resolution on the interstate dynamics, and must therefore be
chosen shorter than the fastest dynamics of interest.

Despite being effective in curing spurious interstate fluctuations, coring of the state trajectory cannot correct for the loss of states in the analysis, because coring is applied after the definition of states.
This means that if the projection step causes the disappearance of free energy barriers (and thus the loss of states), coring of the state trajectory is not
sufficient to recover the full dynamics. 
Recently, it has been shown that missing degrees of freedom can be 
recovered using time-aware representation learning methods exploiting temporal correlations.\cite{wang_latent_2024, diez_recovering_2025}
To introduce a data pre-processing step at the level of the input coordinates, in this work we discuss
`Gaussian filtering'\cite{nagel_selecting_2023, nagel_toward_2023} that
mitigates projection artifacts before clustering, by eliminating the
high-frequency fluctuations in the coordinates trajectory. Acting as a
low-pass filter, it is shown to improve the accuracy of state identification and
subsequent MSM construction. While the use of low-pass filtering has been suggested before in the field of Markov modeling\cite{deng_adaptive_1993,
  jaipal_relative_2017, tan_exploring_2018}, we believe that its impact, utility, and consequences for MSM construction have not been fully appreciated so far.


%
%

\begin{figure}[ht!]
\centering
\includegraphics[width=0.45\textwidth]{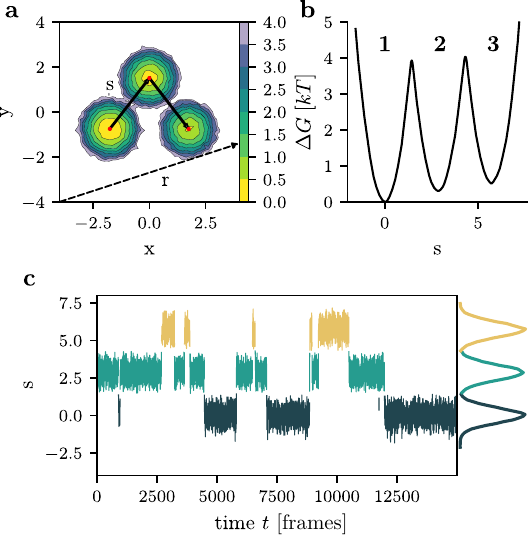}
    \caption{Three-well model. (a) 2D free energy landscape $\Delta G(x, y)$ as a function of the original coordinates $x$ and $y$. Indicated are the optimal 1D coordinate, $s$, directly connecting the three minima, and a suboptimal coordinate, $r$, obtained via the projection on vector $\mbox{\boldmath $r$}$. (b) Projection of the 2D data on the optimal reaction coordinate $s$ yields the barrier-preserving free energy curve $\Delta G(s)$. (c) Trajectory of the optimal reaction coordinate, color coded by the state that each frame is assigned to.}
    \label{fig:toymodel_3states}
\end{figure}

We start with an intuitive example that reveals why artifacts are
introduced by projecting data on a lower dimension and how the
above-mentioned corrective approaches can offer promising solutions.
The model consists of three potential wells separated by barriers of
similar height ($\sim{4}\,k_\text{B}T$), see
Fig.\,\ref{fig:toymodel_3states}. For this simple toy model, it is
straightforward to define a true (i.e., optimal) reaction coordinate,
$s$, following the lines connecting the
minima,\cite{krivov_reaction_2013} i.e., the segment between wells 1
and 2 for $x\leq 0$, and the one between wells 2 and 3 for $x>0$. We
then run an overdamped Langevin simulation to sample the corresponding
2D free energy landscape (Fig.\,\ref{fig:toymodel_3states}a). As shown
by the resulting 1D free energy curve $\Delta G(s)$ in
Fig.\,\ref{fig:toymodel_3states}b, the three minima are clearly
separated along $s$. The frames can be readily assigned to three
states by cutting the curve exactly at the barriers, whose height is
preserved from the 2D case. Figure\,\ref{fig:toymodel_3states}c shows
an example of the time trace of $s$, where the color of each frame
represents the state it is assigned to. All three states are recovered
and clearly separated, because along the true reaction coordinate the
overdamped Langevin trajectory exhibits no immediate recrossing of the
barrier after a transition, as it proceeds to the minimum of the new
state. As a consequence, the state trajectories obtained from the 2D
energy landscape $\Delta G(x, y)$ and the 1D landscape $\Delta G(s)$
are equivalent.

\flushleft
\justifying

The situation becomes more involved in the---more realistic---case in
which we only have a suboptimal reaction coordinate. To mimic this and
illustrate how projection artifacts can arise in our toy model, we
use $s$ as a reference and define in Fig.\,\ref{fig:toymodel_3states}a
two additional 1D reaction coordinates, where the projection effects become progressively worse: the projection on the $x$-axis
$\mbox{\boldmath $x$}=x\,\mbox{\boldmath $e$}_x$, and the projection
on the tilted axis $\mbox{\boldmath $r$} = r\,\mbox{\boldmath $e$}_r$.

\begin{figure*}[ht!]
    \centering
    \includegraphics[width=\textwidth]{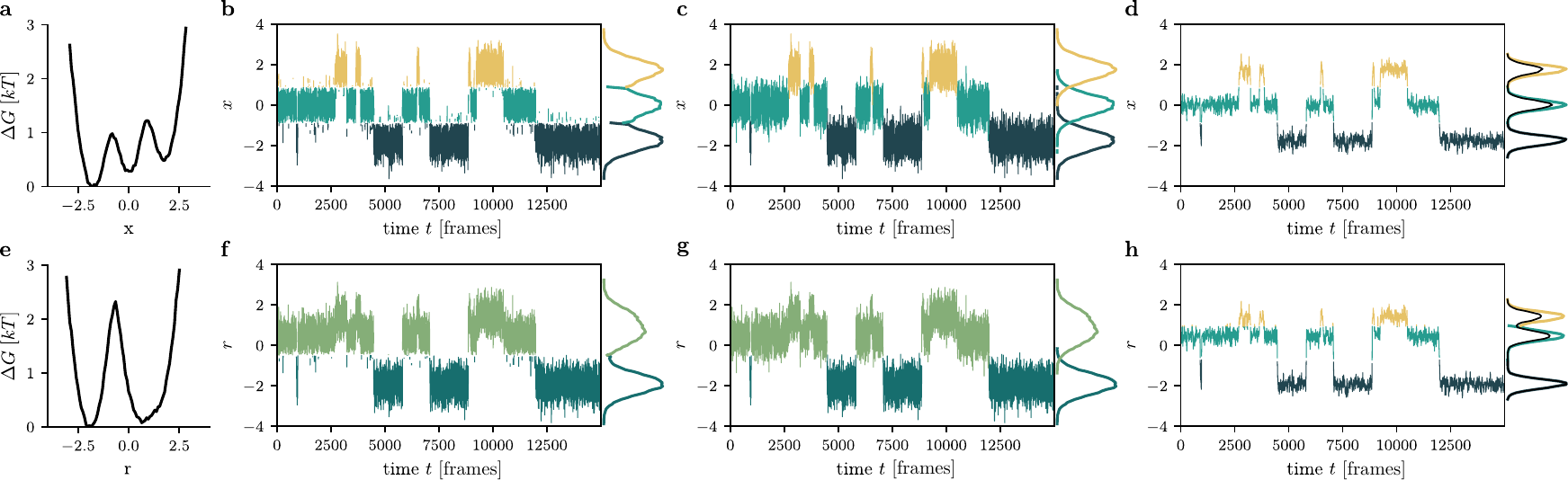}
    \caption{Effects of suboptimal 1D reaction coordinates $x$ (top)
      and $r$ (bottom) chosen for the 2D model. Shown are (left) 1D
      energy curves and (right) color-coded time traces with states
      assigned by cutting at the barrier. Panels (b, f) shows the raw
      data, (c, g) the results after applying iterative coring
      ($t_{\text{cor}}=10\,$frames), and (d, h) the results after
      applying Gaussian filtering ($t_{\text{GF}}=10\,$frames). The right side
      of each panel shows the respective state-resolved
      distributions. }
    \label{fig:toymodel_2states}
\end{figure*}

As shown in Fig.\,\ref{fig:toymodel_2states}a, the three minima of
$\Delta G(x)$ are still clearly 
distinguishable, but the barriers between them become smaller
($\sim{1}\:k_\text{B}T$) than in the original 2D landscape. Moreover, the time
trace of $x$ (Fig.\,\ref{fig:toymodel_2states}b) shows `overshooting'
at the barrier between states: transient fluctuations that are wrongly
identified as transitions between states can be easily spotted as two different colors alternating very rapidly. This happens because frames located at the barriers of $\Delta G(x)$
may already belong to the adjacent state in the orthogonal
$y$–direction. The loss of this information can be seen from the
definition of the 1D free energy as
\begin{equation}
    \Delta G(x) = -k_\text{B}T \ln P(x)
\end{equation} with  \begin{equation}
    P(x) = \int \mathrm{d}y \, P(x,y).
\end{equation} 
When we consider the 2D probability distribution $P(x_{12}, y)$ at the
barrier between states 1 and 2 along $x$ (on the side of state 1), clearly the system
may be already in state 2 along the $y$ coordinate. As a result,
the 1D trajectory $x(t)$ exhibits the typical projection artifact of appearing to fluctuate rapidly back and forth across the barrier.

As a remedy, we apply dynamical
coring,\cite{jain_hierarchical_2014,nagel_dynamical_2019} 
e.g., we request that the trajectory stays at least
$t_{\text{cor}}=10\,$frames in the new state.  
As shown in Fig.\ \ref{fig:toymodel_2states}c, this simple measure
already eliminates most of the spurious transitions at the
barriers. In this way, the coring effectively reduces the overlap of
the states' probability distributions in the border regions.
Since already
small fluctuations may mask true transitions, it is in fact
advantageous to iteratively core the state trajectory,
\cite{diez_markov_2020} by using coring times {incrementing from
  $t_{\text{cor}}=2$ to 10 frames, rather than}
performing a single coring process with
$t_{\text{cor}} = 10\,$frames. Interative coring is
found to reduce the number of reassigned MD frames i.e., it changes
the original trajectory less, while still smoothing with the same time
resolution. \cite{note1}
%

Alternatively, we may apply a Gaussian filter function with standard
deviation $\sigma$ on the feature trajectory $x(t)$,
\cite{nagel_selecting_2023, nagel_toward_2023} 
\begin{equation} \label{eq:GF}
    x(t) \rightarrow \sum_j \frac{1}{\sqrt{2\pi\sigma^2}} \exp \left[ -\frac{(t_j - t)^2}{2\sigma^2} \right] x(t_j).
\end{equation}
It affects a smoothing of $x(t)$ within a moving window of approximately 
$\text{t}_{\rm GF} = 2 \sigma$, corresponding to a low-pass filter with a
cut-off frequency $1/\text{t}_{\rm GF}$. Using $\text{t}_{\rm GF}
=10\,$frames, Fig.\,\ref{fig:toymodel_2states}d reveals that this
smoothing results in a clear separation of the three energy basins.
This is because Gaussian filtering causes a reduction of the amplitude
of the fluctuations, which results in a narrowing of the distance
distributions of the states. Hence we achieve a better resolved
structure of the free energy landscape with clearly defined barriers,
before assigning the time series to states. 

In the second, more drastic, case, we choose $\mbox{\boldmath $r$}= r\,\mbox{\boldmath $e$}_r$
(see Fig.\,\ref{fig:toymodel_3states}a) as projection axis for our
model. The resulting free energy $\Delta G(r)$ (Fig.\
\ref{fig:toymodel_2states}e) reveals that one barrier is lost and only two states can be distinguished.
Figure\,\ref{fig:toymodel_2states}g shows the time-trace $r(t)$ 
after applying iterative dynamical coring ($t_{\text{cor}}=10\,$frames) on
the state trajectory. While the coring eliminates the fast fluctuations
between the two states, it obviously cannot recover the information
that is missing in the free energy landscape $\Delta G(r)$, as
it can only reassign frames among the two 
states that were already identified. Nonetheless, the time evolution of $r(t)$ still suggests the existence of three states (Fig.\,\ref{fig:toymodel_2states}f), hence the question is how to
retrieve this information at the level of the free energy.

\begin{figure}[ht!]
    \centering
    \includegraphics[width=0.45\textwidth]{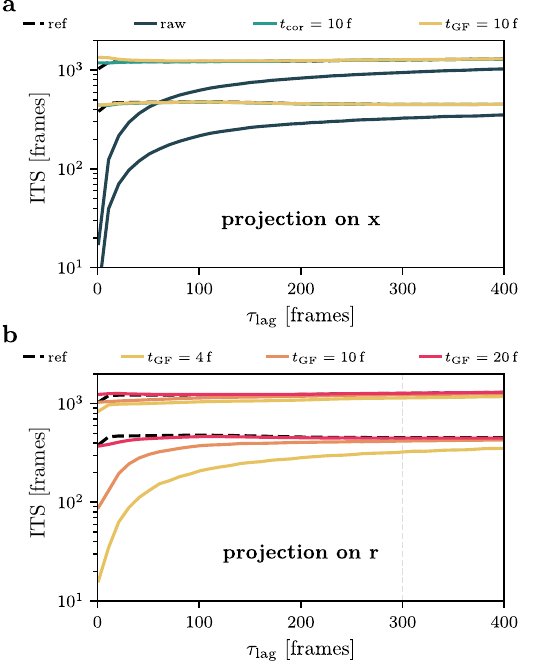}
    \caption{Implied timescales (ITSs) of the toy model, shown as a
      function of the lag time $\tau_{\text{lag}}$.  (a) Projecting on
      coordinate $x$, we show uncorrected data and results from iterative
      coring ($t_{\text{cor}}=10\,$frames, green) and Gaussian
      filtering ($t_{\text{GF}}=10\,$frames, yellow).
      (b) Projecting on coordinate $r$, we use Gaussian filtering
      ($t_{\text{GF}}=4\,, \,10 \,,\,20\, $frames) to
      recover both ITSs. In all cases we compare to the reference
      timescales obtained for the optimal coordinate $s$ (dashed
      black lines).}
        \label{fig:ITS}
\end{figure}

Here the advantage of curing the problem at the level of the input
coordinates is even more obvious, as can be seen in
Fig.\,\ref{fig:toymodel_2states}h. By applying Gaussian filtering to
$r(t)$ with a short window of $t_{\text{GF}}=10\,$frames, the reduced
fluctuations of $r(t)$ facilitate the identification of all three
minima of the free energy $\Delta G(r)$, and therefore allows us to
assign the filtered data to three states. 

To construct a MSM from these data, we compute the transition
probabilities between all states within some lag time
$\tau_{\text{lag}}$. Diagonalization of the resulting transition
matrix yields the implied timescales (ITS) for each value of
$\tau_{\text{lag}}$, which can be used to assess how well an MSM
reproduces the true dynamics of the system,\cite{prinz_markov_2011}
see Fig.\,\ref{fig:ITS}.
%
For the full 2D model, as well as in the case of the optimal reaction
coordinate $s$, we find Markovian behavior, with ITS independent of
the lag time (black dashed lines). The slowest ITS $t_1$ reflects the
main transition between states 1--2 and state 3, while the second
$t_2$ captures the dynamics between state 1--3 and state 2.

Projecting on coordinate $x$, Fig.\,\ref{fig:ITS}a compares the results obtained
for raw data (black), with coring on the state trajectory (green), and with Gaussian filtering on $x$ (yellow). While ITS
obtained from the uncorrected data take a long lag time to converge to the
reference results, both the coring and the Gaussian filtering results
converge rapidly and reproduce the reference perfectly. 

Projecting on coordinate $r$, uncorrected and cored data lead to a
  two-state model and therefore a single ITS. Although coring again
  improves this ITS, it cannot recover the third state.
Gaussian filtering, on the other hand, is found to recover all three
states for a filter window of $t_{\text{GF}} \gtrsim 4\, $frames,
and yields a constant second ITS for
$t_{\text{GF}} = 20\, $frames, see Fig.\,\ref{fig:ITS}b.

Since both the coring time $t_{\text{cor}}$ as well as the
filtering window $t_{\text{GF}}$ introduce a limit on the time
resolution of the model, they need to be chosen shorter than
the fastest timescale of interest. In particular, the lag time of an
MSM built from cored and filtered data should be larger than
$t_{\text{cor}}$ and $t_{\text{GF}}$, respectively. To
preserve a good time resolution of the model, we therefore want to
choose $t_{\text{cor}}$ and $t_{\text{GF}}$ as short as
possible, while still correcting the projection error.
As a heuristic to choose the coring time $t_{\text{cor}}$, Nagel et
al.\cite{nagel_dynamical_2019} suggested to compute the probability
$W_i(t)$ of staying in macrostate $i$ for various coring times, and
choose the shortest $t_{\text{cor}}$ that suppresses the rapid
initial decay reflecting spurious transitions at the barrier. As a
consequence, this procedure yields improved and sufficiently converged
ITSs.
Accordingly, in the case of Gaussian filtering we want to determine
the shortest filtering window $t_{\text{GF}}$ that yields
sufficiently converged ITSs and, in addition, the correct number of
states.
%
%

\begin{figure}[ht!]
	\centering
	\includegraphics[scale=0.9]{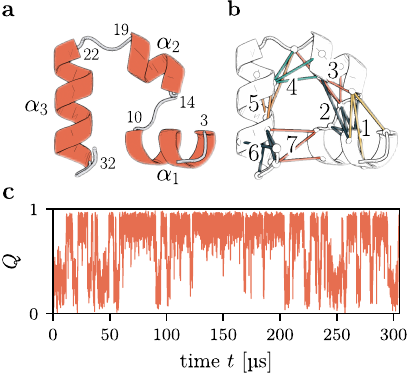}
	\caption{
          The folding of HP35. (a) Structure of the native state, and
          (b) illustration of the structurally most important 27 native 
          contacts, ordered in seven MoSAIC clusters.
          \cite{diez_correlation-based_2022} (c) Time evolution of the fraction of
          native contacts $Q$ obtained from the folding trajectory by
          Piana et al.\cite{piana_protein_2012}. 
          Adapted from Ref.\ \citenum{nagel_toward_2023}.}
	\label{fig:HP35}
\end{figure}

We now study how the virtues of the above presented methods transfer
to the treatment of high-dimensional MD data. To this end, we consider
the folding of villin headpiece \cite{kubelka_chemical_2008}
(Fig.~\ref{fig:HP35}a), for which a \SI{300}{\micro\second}-long MD
trajectory of the fast-folding Lys24Nle/Lys29Nle mutant (HP35) is
publicly available from D.~E.~Shaw Research.\cite{piana_protein_2012}
In a first step we determined the interresidue contacts of
HP35, by assuming a contact to be formed if the distance $d_{ij}$
between the closest non-hydrogen atoms of residues $i$ and $j$ is
shorter than \SI{4.5}{\angstrom}, where $|i-j| > 3$ and $d_{ij}$ is the minimal
distance between all atom pairs of the two
residues.\cite{nagel_toward_2023,yao_establishing_2019} To focus on
native contacts, we request that contacts between these atoms pairs
are populated for more than \SI{30}{\%} of the simulation time, \cite{nagel_toward_2023} which results in 42 native contacts. Using MoSAIC correlation analysis,\cite{diez_correlation-based_2022} the associated contact distances are ordered in clusters of highly correlated contacts. Figure\,\ref{fig:HP35}b shows the 27 contacts of the structurally most important 
clusters, which proceed along the protein backbone from the N- to the C-terminus. As a well-established
1D reaction coordinate,\cite{best_native_2013} we
evaluated along the trajectory the fraction of native contacts formed,
$Q(t)$, which nicely illustrates the time evolution of the
folding process (Fig.~\ref{fig:HP35}c).

To obtain a simple estimate of the folding time, we define the
unfolded region of the free energy landscape by $Q \lesssim 0.3$ and
the native basin by $Q \gtrsim 0.7$, which allows us to directly count the
folding events along the trajectory. Since $Q(t)$ exhibits fast
fluctuations that do not necessarily account for a true folding or
unfolding event, however, we need to invoke a procedure to eliminate
this noise. As for the 2D model studied above, we can either require a
minimum time to stay in the new region (i.e., dynamical coring), or
apply a low-pass filter to the data. In the following, we focus on the
latter and employ Gaussian filtering of the contact distances used in the calculation of $Q(t)$.

\begin{figure}[t]
    \centering
    \includegraphics[width=0.45\textwidth]{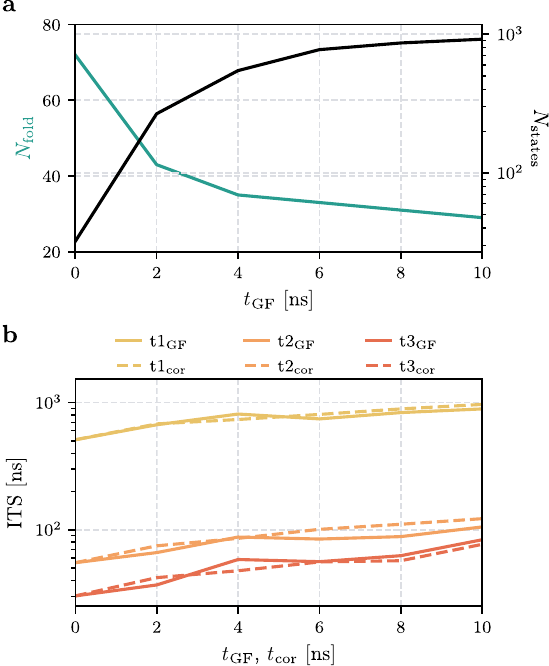}
    
    \caption{(a) Effects of Gaussian filtering with various filtering
      windows $t_{\text{GF}}$ on the observed number of folding events
      $N_{\rm fold}$ of HP35 (green) and on the resulting number of
      microstates $N_{\rm states}$ obtained from robust density-based
      clustering\cite{sittel_robust_2016} (black). (b) First three
      implied timescales of MSMs obtained from coordinates filtered
      with different windows (full lines), and from dynamical coring on the microstates, using different coring
      times (dashed lines).}
    \label{fig:ground_truths_VS_sigma}
\end{figure}

Figure\,\ref{fig:ground_truths_VS_sigma}a shows the effects of Gaussian
filtering on $Q(t)$, by depicting the number of folding events as a
function of the filtering window $t_{\text{GF}}$ applied to the
contacts. While the uncorrected data overestimate the number of
folding events by a factor 2, we find that it takes
$t_{\text{GF}} \gtrsim 4\,$ns to converge to about 30 folding
events. This corresponds to a folding time of
$\sim \SI{2}{\micro\second}$, which is in agreement with previous
works. \cite{piana_protein_2012} Yielding the correct folding time
with minimal smoothing, $t_{\text{GF}} = 4\,$ns was also chosen
in the benchmark study by Nagel et al.\cite{nagel_toward_2023}

As discussed above, a major advantage of
Gaussian filtering is that it can identify hidden
conformational states, resulting in an improved MSM.
To explore this, we use the 42 native contact distances of HP35 as
input coordinates in the subsequent analysis, and apply 
Gaussian filtering directly on these features.
 As dimensionality reduction step we use principal component analysis
on the smoothed contact distances. \cite{ernst_contact-_2015} The first five
components exhibit a multimodal structure of their free energy curves,
reveal the slowest timescales ($\sim 0.1$ to $\SI{2}{\micro\second}$),
and explain $\sim \SI{80}{\%}$ of the total correlation.\cite{nagel_toward_2023}
Using these collective variables, we next perform robust density-based
clustering.\cite{sittel_robust_2016} This clustering algorithm  computes a local free
energy estimate for every frame of the trajectory, by counting all
other structures residing in a hypersphere of fixed radius. Reordering
all structures by increasing free energy, the method directly
yields the minima of the free energy landscape, where the number of
these microstates depends on the chosen minimal population of a state
(here $P_\text{min}=\SI{0.01}{\%}$). (This is unlike
the popular $k$-means clustering, where this number needs to be
specified beforehand.)

Using different windows $t_{\text{GF}}$ to filter the contact distances, the
above workflow yields clusterings with different number of microstates $N_{\rm states}$. Remarkably, Fig.\,\ref{fig:ground_truths_VS_sigma}a
shows that $N_{\rm states}$ increases drastically with $t_{\text{GF}}$, from
32 states for the uncorrected data ($t_{\text{GF}} = 0$), via 547 states
($t_{\text{GF}} = 4\,$ns) to 990 states ($t_{\text{GF}} = 10\,$ns), and stays
approximately constant thereafter.  This indicates that Gaussian
filtering provides a significantly higher resolution of the underlying
free energy landscape, confirming what we showed earlier for our toy
model example.

\begin{figure*}[ht!]
    \centering
    \includegraphics[width=\textwidth]{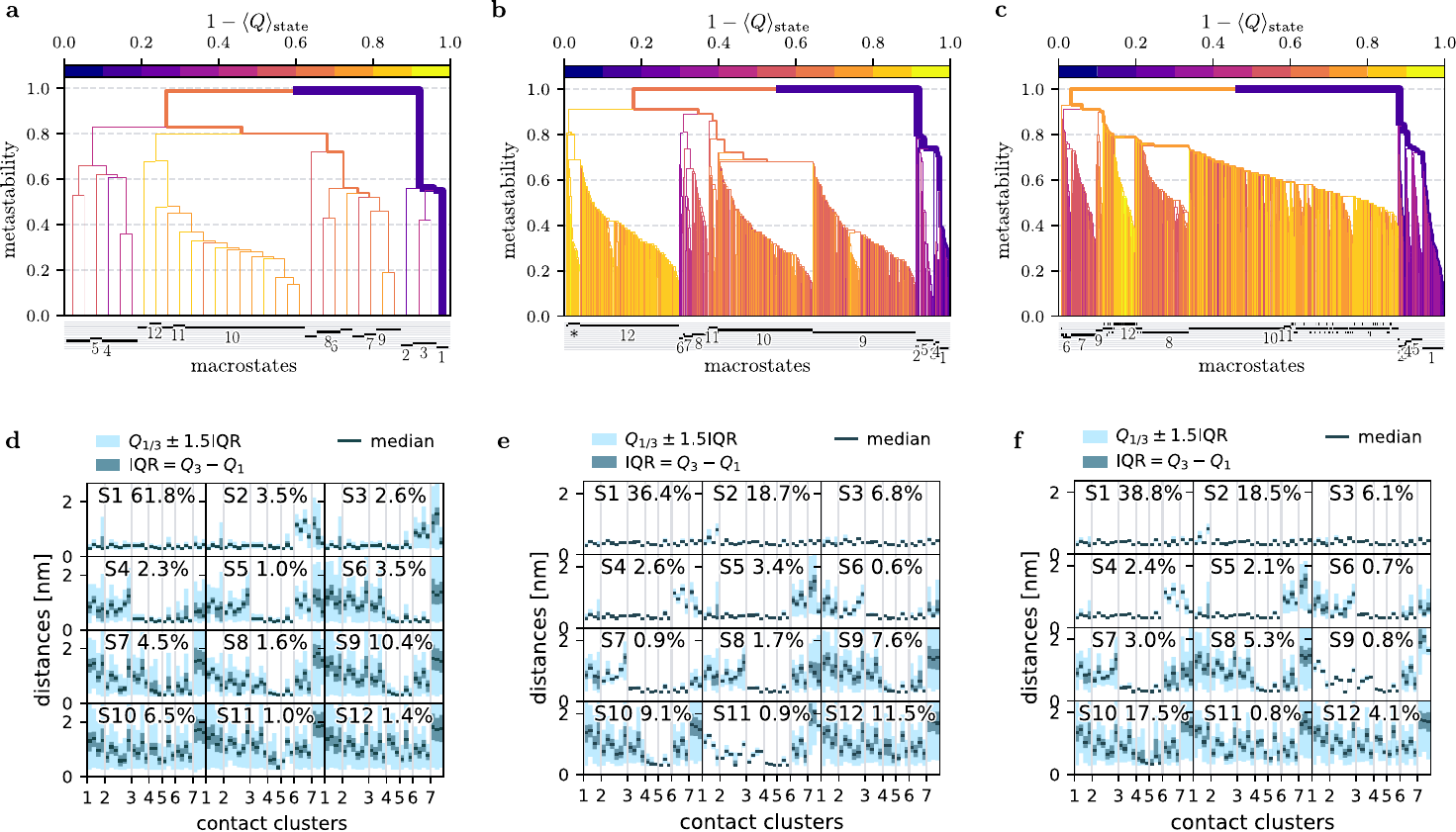}
    \caption{Construction of metastable conformational states
      describing the folding of HP35. Compared are uncorrected (a)
      data and Gaussian filtered data using (b) $t_{\text{GF}}=4\,$ns and (c)
      $t_{\text{GF}}=10\,$ns. (Top) MPP dendrogram showing the hierarchical
      clustering of microstates into twelve metastable macrostates,
      using a lag time of 10\,ns. Colors indicate the fraction of
      native contacts $Q$. (Bottom) Structural characterization of the
      resulting twelve metastable states of HP35, ordered by
      decreasing $Q$. For each state, the distribution of contact
      distances are represented by the median $Q_2$, the interquartile
      range $\text{IQR}=Q_3- Q_1$ and the lower (upper) bound as the
      smallest (largest) data point in
      $Q_{1/3} \pm 1.5\cdot\text{IQR}$}
    \label{fig:dendro_HP35}
\end{figure*}

To allow for a simple interpretation of an MSM, the numerous
microstates are lumped into a few macrostates, which concisely account
for the considered dynamical process. In this framework, differences
between microstate partitionings do matter, because they may result in
different macrostates and henceforth different dynamics. To group
microstates into macrostates, we use the most probable path (MPP)
algorithm, \cite{jain_identifying_2012, jain_hierarchical_2014} which
constructs the transition matrix of the microstates and successively
merges the state with the lowest metastability into the one to which
it has the highest transition probability. Iteratively repeating this
procedure yields a dendrogram that illustrates how the microstates
aggregate into larger energy basins, which correspond to metastable
conformational states.  To resolve at least the first tier of the
emerging hierarchical structure, we rely on the improved MPP
algorithm\cite{nagel_selecting_2023} that ensures that the states have
a minimum metastability of 0.5 and a minimum population of 0.5\,\%,
thus obtaining 12 metastable states.
Showing these dendrograms for (a) uncorrected and (b,c)
Gaussian-filtered data with $t_{\text{GF}} = 4\,$ns and 10\,ns,
Fig.\,\ref{fig:dendro_HP35} clearly reveals the much larger number of
microstates identified with Gaussian filtering. This leads to distinctly
different lumping patterns 
and therefore also to different metastable states.

To illustrate this finding, Figs.\,\ref{fig:dendro_HP35}d-f present structural characterizations of the three resulting macrostate
partitionings. For each state, the distributions of contact
distances in MoSAIC  clusters 1--7 (Fig.\,\ref{fig:HP35}b) are shown.
The 12 macrostates are ordered by decreasing fractions of native
contacts formed, such that state 1 corresponds to the native state
(all distances shorter than $\SI{4.5}{\angstrom}$), whereas state 12
represents the fully unfolded state, with a broad distribution of
large distances.

We first discuss the Gaussian-filtered data with $t_{\text{GF}}=4\,$ns (panel
e) as used in the benchmark study by Nagel et
al.,\cite{nagel_toward_2023} and are here referred to as `reference
results'. The first three states are structurally well-defined,
native-like conformations that differ mainly in the details of helix
1. States 4 and 5, which exhibit broken contacts near the C-terminus,
still belong to the native energy basin. States 6–8 are sparsely
populated ($\lesssim \SI{1}{\%}$) intermediates, while the unfolded
basin comprises states 9–12, showing an increasing degree of
structural disorder. 
In particular, we find a completely unfolded state 12 and two
largely unfolded states 9 and 10 (with mainly intact contacts between
helices $\alpha_2$ and  $\alpha_3$, see Fig.~\ref{fig:HP35}b), which 
renders transitions from state 12 to states 9 and 10 the most likely
initial step of the folding process.\cite{nagel_selecting_2023}

In contrast, the contact analysis of the uncorrected data (panel d)
suggests that the first three states are merged into a single
state. Obviously, the suppression of high-frequency fluctuations in
the input data resulted in a better structural resolution of the
native basin of HP35. Furthermore, the six states corresponding to the
unfolded basin appear more congested and less distinctly separated
than those obtained from the filtered reference data. 

On the other hand, it is interesting to check if the state
partitioning obtained for $t_{\text{GF}}=10\,$ns (where the number of
microstates is maximal) presents an improvement over the
reference results with $t_{\text{GF}}=4\,$ns. While the first six states are
virtually the same, we find that states 7 and 8 of the reference are
lumped in one state, and that reference state 11 turns into state 9. 
The remaining unfolded states look similar to the reference state, but
they do not clearly discriminate between the completely unfolded state
and largely unfolded states with intact contacts between $\alpha_2$
and $\alpha_3$.

Apart from the structural properties of the macrostates, it is
important to assess the dynamics of the various MSMs. To this end,
Fig.\,\ref{fig:ground_truths_VS_sigma}b shows their first three
implied timescales (ITSs) as a function of the filtering window
$t_{\text{GF}}$, assuming a lag time of 10\,ns. We find a clear increase of
the first ITS when comparing unfiltered ($t_{\text{GF}}=0$) and reference
($t_{\text{GF}}=4$\,ns) results and a modest further improvement when using
$t_{\text{GF}}=10$\,ns.
Hence Gaussian filtering improves the Markovianity of the model, by
increasing the energy barriers due to a higher resolution of the free
energy landscape.

Following the discussion of the 2D model,
we therefore conclude that $t_{\text{GF}}=4\,$ns used in the benchmark study
by Nagel et al.\cite{nagel_toward_2023} represents a sweet spot of the
filtering window range. As shown in Fig.\,\ref{fig:ground_truths_VS_sigma},
this choice yields good ITSs, the correct number of folding events
$N_{\rm fold}$, and a significantly increased number of microstates
$N_{\rm states}$. (Note that changing $t_{\text{GF}}$ from 0 to 4\,ns results
in an increase of $N_{\rm states}$ by $\sim 1600\%$, while changing 
to 5\,ns only gives a further increase of $\sim 80\%$.) Resulting in a
time resolution of $t_{\text{GF}}=4\,$ns,
this minimal size of the filtering window still allows using relatively
short lag times in the construction of the MSM.

We finally compare the above MSMs obtained using Gaussian filtering of the input coordinates to MSMs obtained using iterative dynamical coring of the
microstates. To this end, Fig.\,\ref{fig:ground_truths_VS_sigma}b
shows the first three ITSs as a function of the coring time
$t_{\text{cor}}$. Recalling that we need to choose
$t_{\text{cor}} \approx t_{\text{GF}}$ to achieve a similar time
resolution, we find that coring and filtering yield virtually the
same timescales. However, as coring is applied to the microstates,
it cannot achieve the improved structural resolution of the free
energy landscape found for filtering
(see Fig.\,\ref{fig:dendro_HP35}). In fact, the structural
characterization of the macrostates obtained from coring (with
$t_{\text{cor}}=4\,$ns) is virtually identical to the results
obtained without coring shown in Fig.\,\ref{fig:dendro_HP35}d.
In other words, dynamical coring only improves the dynamics of the
resulting MSM, while Gaussian filtering improves in addition the
structural properties of the macrostates of the model. Due to the
complementary nature of the two methods, it may nevertheless be
advantageous to use both corrections in order to optimize
the Markovianity of the model.

%
%

In conclusion, we have shown that Gaussian filtering of input
coordinates can significantly improve the performance of an MSM. Apart
from naturally improving the timescale separation and therefore the
implied timescales of the model, the reduction of the fluctuations may
greatly facilitate the identification of microstates in the subsequent
clustering of the data.
We emphasize that this improved representation of the free energy
landscape is only achieved because the filtering is applied at the
very beginning of the trajectory analysis. In contrast, related
approaches that correct for projection artifacts—such as dynamical
coring of the state trajectory—cannot provide the same
benefit. As the method is fully compatible with any subsequent
workflow used to construct an MSM, we propose that initial
Gaussian filtering of the feature trajectory should become a standard
component of the MSM workflow. 


\subsection*{Acknowledgments}

This work has been supported by the Deutsche
Forschungsgemeinschaft (DFG) within the framework of the Research Unit
FOR 5099 “Reducing complexity of nonequilibrium” (project
No. 431945604), the High Performance and Cloud Computing Group at the
Zentrum für Datenverarbeitung of the University of Tübingen and the
Rechenzentrum of the University of Freiburg, the state of
Baden-Württemberg through bwHPC and the DFG through grant no INST
37/935-1 FUGG (RV bw16I016), and the Black Forest Grid Initiative.

\subsection*{Data Availability Statement}

The simulation data and all intermediate results for our reference
model of HP35, including our software packages \emph{MoSAIC},\cite{diez_correlation-based_2022}
\emph{FastPCA},\cite{sittel_principal_2017} \emph{Clustering}\cite{sittel_robust_2016} and
\emph{msmhelper}\cite{nagel_msmhelper_2023} and detailed descriptions to reproduce all steps of the workflow, can be downloaded from
https://github.com/moldyn/HP35. 

\bibliography{\dir/StockZotero21.1.2026,\dir/new}

\begin{mcitethebibliography}{44}
\bibitem[not()]{note1}
{The improvement is modest for the simple 2D model: for $\text{t}_\text{cor} =
  10\,$frames standard and iterative coring reassign $5.14\%$ vs.\ $5.33\%$ of
  the MD frames, respectively; for $\text{t}_\text{cor} = 50\,$frames, though,
  we already find $5.29\%$ vs.\ $13.95\%$. For a complex system such as HP35
  considered below the effect is more substantial already at short coring
  times, e.g., we obtain $7.79\%$ vs.\ $11.57\%$ for $\text{t}_\text{cor} =
  2\,$ns, and $8.98\%$ vs.\ $15.03\%$ for $\text{t}_\text{cor} = 4\,$ns}\relax
\mciteBstWouldAddEndPuncttrue
\mciteSetBstMidEndSepPunct{\mcitedefaultmidpunct}
{\mcitedefaultendpunct}{\mcitedefaultseppunct}\relax
\EndOfBibitem
\end{mcitethebibliography}

\end{document}